\documentclass[global,twocolumn]{svjour}
\usepackage{graphics}
\journalname{Applied Physics A}

\newcommand{\Sarah}{SARA{\it h} }

\begin{document}
\title{Symmetry and magnetic structure determination: \newline Developments in refinement techniques and examples}
\author{A.~S. Wills\inst{1} 
}                     
\offprints{wills@ill.fr}          
\institute{Institut Laue-Langevin, 6 Rue Jules Horowitz, BP 156, 38042 Grenoble Cedex 9, France.}

\date{Received: date / Revised version: date}
%
\maketitle
\begin{abstract}

Group Theory techniques can aid greatly the determination of magnetic structures. The integration of their calculations into new and existing refinement programs is an ongoing development that will simplify and make more rigorous the analysis of experimental data. This paper presents an overview of the practical application of symmetry analysis to the determination of magnetic structures. Details are given of the different programs that perform these calculations and how refinements can be carried out using their results. Examples are presented that show how such analysis can be important in the interpretation of magnetic diffraction data, and to our reasoning of the causes for the observed ordering.

\end{abstract}


\section{Introduction}
\label{intro}

Despite immense technical progress since the first magnetic neutron diffration experiments of Shull and Smart,\cite{Shull_Smart} the determination of magnetic structures remains a subject that is typically limited by the data analysis strategy: structures are generally determined by intuition or simple trial and error refinement. As a consequence, the literature is full of incorrect magnetic structures and incomplete refinements. While Group Theory techniques can be applied to limit the number of trial structures, or to determine along which directions the spin components can lie, their calculations when carried out by hand are arduous. This has led to their being applied only when a problem warrents their use, or more commonly, when it is sufficiently close to an example already in the literature. Recently, a number of computer programs have been developed that  allows the unspecialised user to perform these calculations automatically. Their integration with common refinement codes allow for the first time the simple and rigorous examination of which symmetry-allowed magnetic structures are compatible with collected data.

In this article a brief overview is made of the practical application of using Group Theory to aid the determination of magnetic structures from both powder and single crystal samples. Examples are given that demonstrate the importance of symmetry information for the correct analysis of magnetic diffraction data, and concommitently to the understanding of the physical reasons for the formation of a long-ranged magnetic order.

\section{Representational Analysis Calculations}

The calculations can be separated into two parts. The first is a grouping according to symmetry of the possible magnetic structures that are compatible both with the Space Group of the crystal structure and the propagation vector {\bf k} of the magnetic ordering. The second part involves the application of Landau theory as a tool to simplify which of these are possible as a result of a continuous second-order phase transition. While the grouping and the labelling of the different magnetic structures by their symmetry properties is completely general, the assumptions made that involve the Landau theory are subject to its limitations.

The application of Group Theory to the determination of magnetic structures is termed {\it Representational Analysis}\cite{Bertaut1,Bertaut2,Bertaut3,Bertaut4,Izyumov-book} and is based on the calculation of the Fourier components of an  ordered magnetic structure that are compatible with the symmetry of the crystal space group before the phase transition and the propagation vector.

The first step in the analysis is the identification of the propagation vector {\bf k} associated with the phase transition, and which space group symmetry operations leave it invariant. These operations form the little group G$_{\bf k}$. The symmetry elements of G$_{\bf k}$ and value of {\bf k} are then used to determine the different Irreducible Representations (IRs) of G$_{\bf k}$. The different basis vectors (Fourier components of the magnetic structure),BV, that are projected out from an Irreducible Representation define a basis vector space that may be termed a ``symmetry-allowed'' model. The different IRs define orthogonal basis vector spaces that can be used to conveniently classify the different possible magnetic structures.

\section{Application of Landau theory}

The Landau theory of a second-order phase transition requires that the Hamiltonian of the system is invariant under the symmetry operations of G$_{\bf k}$. This leads to the requirement that for a second-order phase transition an ordered structure can be the result of only a {\it single} IR becoming critical. This typically leads reduces the number of trial structures and the variables that each involve.

When the assumptions of Landau Theory are not valid, for instance when the Hamiltonian possesses odd powered terms the mixing of components from different IRs becomes possible. Continuing with this logic, the observation of a magnetic structure that involves different IRs is suggestive of either successive ordering transitions for each IR, or additional terms in the magnetic Hamiltonian that relax the single IR rule, e.g. Crystal Field terms at sites with certain point symmetries.

\section{Programs that perform these calculations}
\label{programs}

While tabulated values of the IRs of the Space Groups have existed for many years,\cite{Bradley_Cracknell,Kovalev1} they are prone to the inaccuracies\footnote{the exception to this is likely to be those presented in Ref. \cite{Kovalev} as these have been subject to computer verification.} and their laborious use in calculations carried out by hand is perhaps the major reason for their restricted use. Preferable to many are the programs that have been written that calculate these IRs, or use files of tabulated values such as: KAREP\cite{Karep} (calculated), MODY\cite{Mody} (unverified tabulated values), BasiReps\cite{Basireps} (based on KAREP), and \Sarah\cite{Wills-Sarah} (with the choice of KAREP-based and computer verified tabulated values). MODY, BasiReps, and \Sarah also use these values to calculate directly the basis vectors associated with the different IRs, and so allow the rapid and simple calculation of the possible symmetry-allowed structures.

\section{Refinement using the results of symmetry analysis}

The simplest and most general mathematical description of a magnetic structure is in terms of Fourier components: the basis vectors that result from the Group Theory.
The panoply of different possible commensurate and modulated incommensurate structures can be simply understood in terms of the form of the basis vector(s) for a site and the value(s) of {\bf k}. To simplify the refinement process, SARA{\it h} and BasiReps have been written to integrate with the standard refinement codes FullProf 2000\cite{Fullprof} (BasiReps and SARA{\it h}) and GSAS\cite{GSAS} (SARA{\it h}). The applicability of the technique is now limited principally by the choice of refinement codes\footnote{Perhaps the most important restriction arises from the absence of a propagation vector in GSAS- it is consequently limited to only simple commensurate structures.}.

The reduction in data due to the magnetic form factor and the observation of only the component of the magnetisation perpendicular to the scattering vector result in instabilities and limitations when using conventional least-squares refinements. These can be overcome by the reverse-Monte Carlo based algorithms which allow the automatic exploration of the degrees of freedom associated with a given magnetic structure and the identification of additional minima in the refinement\cite{Wills-Sarah}. In more complex cases the technique of Simulated Annealing can be employed: this uses a decreasing criterium to allow the more controlled evolution of the system that is required to bypasses the false minima commonly associated with larger numbers of variables.

\section{On-going development of refinement codes}

Due to the difficulties in magnetic structure refinement being greatest for data from powders, development of these new magnetic structure refinement codes has concentrated on their analysis. As part of a collaboration between the Insitut Laue-Langevin, the Commissariat \`a l'\'Energie Atomique, and the Laboratire L\'eon Brillouin we are at present working on the extension of the FullProf package not only towards the refinement of unpolarised and polarised neutron diffraction data collected from single crystals, but also to data collected by the technique of spherical neutron polarimetry. After these developments the data collected by any technique, from conventional powder diffraction to even the most complex collection techniques, will be refinable in terms of symmetry generated basis vectors.

\section{Canted antiferromagnetism in M$_2$[Ni(CN$_2$)] (where M=Mn and Fe)}

The first example of the application of these techniques is taken from the M$_2$[Ni{CN$_2$)] (where M=Mn and Fe) molecular solids.\cite{Lappas} These binary metal-dicyanide molecular materials crystalise in the space group $P$nnm. Below second order transitions at $T\sim 16$~K for the Mn and $T\sim 19$~K  for the Fe compound, these display long-range magnetic order with the propagation vector {\bf k} = (000). Symmetry analysis of the magnetic M atom at the 2a position indicates that there are 4 symmetry-allowed models; these correspond to the IRs $\Gamma_1$, $\Gamma_3$, $\Gamma_5$, and $\Gamma_7$ in the notation of Kovalev\cite{Kovalev1,Kovalev} and their basis vectors are given in Table \ref{Table1}. While data collected using powder neutron diffraction can be well fitted by a simple model of antiparallel spins (M1=-M2) that were free to rotate in the $ab$ plane, this spin structure is {\it not} allowed by symmetry. Investigation of the symmetry-allowed models found that the data could only be well fitted by $\Gamma_5$.

\begin{table}[h]
\begin{tabular}{cccc}
\hline
 IR & BV & M1 & M2 \\
\hline 
$\Gamma_1$ & $\psi_1$  & (0 0 1) & (0 0 $\bar{1}$) \\
$\Gamma_3$ & $\psi_2$  & (1 0 0) & (1 0 0) \\
  & $\psi_3$  & (0 1 0) & (0 $\bar{1}$ 0) \\
$\Gamma_5$ & $\psi_4$  & (1 0 0) & ($\bar{1}$ 0 0) \\
  & $\psi_5$  & (0 1 0) & (0 1 0) \\
$\Gamma_7$ & $\psi_6$  & (0 0 1) & (0 0 1) \\
\hline
\end{tabular}
\caption{Magnetic basis vectors, BVs, for the 2a site of the space group {\it P}nmm with the propagation vector {\bf k} = (0 0 0). M1 and M2 are the atomic positions (0 0 0) and ($\frac{1}{2}$ $\frac{1}{2}$ $\frac{1}{2}$).}
\label{Table1}
\end{table}

Inspection of the associated basis vectors ($\psi_4$ and $\psi_5$) shows that while the moments are antiferromagnetically aligned along $a$, an uncompensated magnetisation can exist along $b$. The presence of such a ferromagnetic spin canting has been confirmed by  dc susceptibility data.

\section{Rare earth nickel borocarbides}

Very recently symmetry analysis has provided important new information about the variety of magnetic orderings that are observed in the rare-earth nickel borocarbide $R\rm Ni_2B_2C$ ($R$ = Gd--Lu, Y).\cite{Wills_Canfield1} In these materials Fermi surface nesting effects propagated via the RKKY interactions create a strong tendancy for these materials to order magnetically with the propagation vector {\bf k}=(0.55 0 0). Despite this, a large number of different magnetic structures are observed for the series. 

The key to understanding the magnetism of these materials was the single-ion anisotropies of the rare-earths. These are typically well defined and possess a characteristic energy scale that is far greater than that of the exchange interactions. Their effect is to force the magnetisation to point along specific crystalographic directions. Symmetry analysis of the different magnetic structures that are possible for the propagation vector {\bf k}=(0.55 0 0) in this system indicated that when the single-ion effects were incompatible with the symmetry-allowed directions for this propagation vector, the system orders according to another propagation vector that {\it does} allow the single-ion effects of the rare-earth in question to be satisfied.

\section{The jarosites }

The jarosites (AFe$_3$(SO$_4$)$_2$(OD)$_6$ where A=Na$^+$, K$^+$, Rb$^+$, Ag$^+$, ND$_4^+$, $\frac{1}{2}$Pb$^{+2}$) are the most studied examples of {\it kagom\'e} antiferromagnets. They have been the objects of much scrutiny as the magnetic sublattice makes up a geometry of vertex sharing triangles. This results in their having an infinite number of classical ground states in the presence of only nearest-neighbour antiferromagnetic exchange interactions. Further-neighbour interactions can raise this degeneracy and cause one particular ordered spin configuration to be favoured from the degenerate manifold. Unfortunately, the experimental determination of which occurs is strongly hindered by an ignorance of the particular degeneracy breaking interaction. Not only did symmetry analysis provide a particularly effective tool for the reduction in the number of trial structures, but it also helped understood them in terms of the different terms in the exchange Hamiltonian.\cite{Wills-PRB-Jarosites}

\section{Conclusion}

The tools are now available that allow the unspecialised researcher to use symmetry analysis to make simpler and more rigorous the refinement of magnetic neutron diffraction data. As demonstrated here, their application to even the simplest structures is important as physically unreasonable models still often fit experimental data.


\begin{thebibliography}{}
%
\bibitem{Shull_Smart}
C.~G. Shull and J.~S. Smart, Phys. Rev. {\bf 76}, 1256 (1949).

\bibitem{Bertaut1}
E.~F. Bertaut, J. Appl. Phys. {\bf 33}(1962),1138.

\bibitem{Bertaut2}
E.~F. Bertaut, Acta. Cryst. A{\bf 24} (1968), 217.

\bibitem{Bertaut3}
E.~F. Bertaut, J. de Physique Colloque, {\bf C1} (1971), 462.

\bibitem{Bertaut4}
E.~F. Bertaut, J. Magn. Magn. Mat.{\bf 24} (1981), 267.

\bibitem{Izyumov-book}
Yu.~A. Izymov, V.~E. Naish, and R.~P. Ozerov, {\it Neutron Diffraction
of Magnetic Materials} (Consultants Bureau, New York 1991).

\bibitem{Bradley_Cracknell}
C.~J. Bradley and A.~P. Cracknell, {\it The Mathematical Theory of Symmetry in Solids} (Clarendon Press, Oxford 1972)

\bibitem{Kovalev1}
O.~V. Kovalev, {\it Irreducible Representations of the Space Groups} (Gordon and Breach, New York 1965).

\bibitem{Kovalev}
O.~V. Kovalev, {\it Representations of the Crystallographic Space
Groups} Edition 2 (Gordon and Breach Science Publishers, Switzerland
1993).

\bibitem{Karep}
E.~R. Hovestreydt, M.~I. Aroyo and H. Wondratschek, J. Appl. Cryst. {\bf 25} (1992), 544.

\bibitem{Mody}
P. Czapnik and W. Sikora, program available from www.ftj.agh.edu.pl/zffs/staff/sikora/program/mody.zip. 

\bibitem{Basireps}
J. Rodr\'{\i}guez-Carvajal, program available from juan@llb.saclay.cea.fr.

\bibitem{Wills-Sarah}
A.~S. Wills, Physica B {\bf 276} (2000), 680. Program available
from ftp://ftp.ill.fr/pub/dif/sarah/

\bibitem{Fullprof}
J. Rodr\'{\i}guez-Carvajal, Physica B {\bf 192} (1993), 55.

\bibitem{GSAS}
A.~C. Larsen and R.~B. von Dreele, program available from 
ftp://ftp.lanl.gov/public/gsas/

\bibitem{Lappas}
A. Lappas, M. Kurmoo, and K. Prassides, J. Phys.: Condens. Matter. (2001), {\it submitted}.

\bibitem{Wills_Canfield1}
A.~S. Wills, C. Detlefs, and P.~C. Canfield, Eur. Phys. J. B (2001), {\it submitted}.

\bibitem{Wills-PRB-Jarosites}
A.~S. Wills, Phys. Rev. B, {\bf 63} (2001), 064430.

\end{thebibliography}
\end{document}